\documentclass[conference]{IEEEtran}
\IEEEoverridecommandlockouts
\usepackage{cite}
\usepackage{amsmath,amssymb,amsfonts}
\usepackage{algorithm}
\usepackage{algpseudocode}
\usepackage{graphicx}
\usepackage{textcomp}
\usepackage{xcolor}
\usepackage{soul}
\usepackage{enumitem}
\usepackage{pifont}

\usepackage{flushend}

\usepackage{booktabs}
\usepackage{multirow}
\usepackage{tabularx}
\usepackage{diagbox}
\usepackage{colortbl}

\usepackage{makecell}

\usepackage[caption=false]{subfig}
\usepackage[font=footnotesize,labelfont=bf]{caption}

\def\BibTeX{{\rm B\kern-.05em{\sc i\kern-.025em b}\kern-.08em
    T\kern-.1667em\lower.7ex\hbox{E}\kern-.125emX}}

\usepackage[absolute]{textpos}

\begin{document}
\begin{textblock}{5}(11.8,0.6)
(Special Session)
\end{textblock}

\begin{textblock}{14}(5.7,0.8)
This paper will be presented at IEEE VLSI Test Symposium (VTS) 2025.
\end{textblock}

\IEEEoverridecommandlockouts
\IEEEpubid{\makebox[\columnwidth]{ 979-8-3315-2144-8/25/\$31.00 \copyright2025 IEEE \hfill} \hspace{\columnsep}\makebox[\columnwidth]{ }}

\title{LLM-IFT: LLM-Powered Information Flow Tracking for Secure Hardware\\
}

\author{\IEEEauthorblockN{Nowfel Mashnoor, Mohammad Akyash, Hadi Kamali, Kimia Azar}
\IEEEauthorblockA{\textit{Department of Electrical and Computer Engineering (ECE), University of Central Florida, Orlando, FL 32816, USA} \\
\{nowfel.mashnoor, mohammad.akyash, kamali, azar\}@ucf.edu}
}

\maketitle

\begin{abstract}
As modern hardware designs grow in complexity and size, ensuring security across the confidentiality, integrity, and availability (CIA) triad becomes increasingly challenging. Information flow tracking (IFT) is a widely-used approach to tracing data propagation, identifying unauthorized activities that may compromise confidentiality or/and integrity in hardware. However, traditional IFT methods struggle with scalability and adaptability, particularly in high-density and interconnected architectures, leading to tracing bottlenecks that limit applicability in large-scale hardware. To address these limitations and show the potential of transformer-based models in integrated circuit (IC) design, this paper introduces LLM-IFT that integrates large language models (LLM) for the realization of the IFT process in hardware. LLM-IFT exploits LLM-driven structured reasoning to perform hierarchical dependency analysis, systematically breaking down even the most complex designs. Through a multi-step LLM invocation, the framework analyzes both intra-module and inter-module dependencies, enabling comprehensive IFT assessment. By focusing on a set of Trust-Hub vulnerability test cases at both the IP level and the SoC level, our experiments demonstrate a 100\% success rate in accurate IFT analysis for confidentiality and integrity checks in hardware. 
\end{abstract}


\section{Introduction}

As modern hardware designs become increasingly complex, security concerns are growing, especially in system-on-chip (SoC) that integrate multiple intellectual property (IP) modules from different entities \cite{tehranipoor2024hardware, hu2020overview}. Almost all hardware designs now rely on third-party IP (3PIP) cores, electronic design automation (EDA) tools, and globalized supply chain (e.g., outsourced manufacturing) to reduce time-to-market (TTM) and costs. However, due to the lack of transparency within parties, these practices introduce security threats, e.g., unauthorized data access and design malicious modification. 

As hardware designs grow more complex and globalized, ensuring security with formal guarantees has become a critical challenge \cite{farzana2019soc, azar2022fuzz}. Traditional formal verification methods, semi-formal testing methods, and even machine learning (ML)-based approaches, while effective, are often costly and time-consuming, particularly for large-scale SoCs\footnote{Formal methods, e.g., model checking and assertion-based verification, require cycle-accurate analysis and experts' knowledge and effort \cite{witharana2022survey} .}\footnote{Testing technique, e.g., fuzz testing requires security-oriented mutation and feedback analysis, making them less applicable at SoC level \cite{jain2021survey, hossain2024fuzzing}.}\footnote{ML-based approaches requires extensive labeled datasets and struggle with generalization across different hardware designs \cite{fan2024efficient, yasaei2022hardware}}. Among existing approaches, information flow tracking (IFT) is a more efficient approach to verifying security properties like confidentiality and integrity \cite{hu2021hardware}. By analyzing data propagation within hardware (by labeling and tracking data movement), IFT helps identify vulnerabilities from design flaws, timing side channels, and malicious modifications like Hardware Trojans (HT), making it vital for secure hardware design \cite{tai2020multi}. 

While Information Flow Tracking (IFT) methods are systematically beneficial, they have notable limitations. Static IFT often struggles with scalability in complex designs and often over-approximates, leading to false positives \cite{nahiyan2017hardware}. Moreover, static IFT lacks the capability to quantify the severity of information leakage \cite{tai2020multi, reimann2021qflow}. Dynamic IFT, although better in scalability, relies heavily on test coverage, limiting its effectiveness in large-scale security verification and potentially missing certain vulnerabilities. \cite{solt2022cellift}. Additionally, many IFT techniques rely on manually crafted taint rules, making them difficult to generalize across different hardware architectures \cite{solt2022cellift, solt2024hybridift}. These constraints reduce their adaptability to emerging threats and complex security violations. Addressing these challenges requires more scalable and adaptable IFT to secure modern hardware systems effectively.

To address these challenges, we introduce LLM-IFT, the first-of-its-kind IFT that leverages large language model (LLM) for hardware security verification. LLM-IFT relies on LLM-driven reasoning over structured data (e.g., graphs \cite{jin2024large}) and consists of four main steps: \ul{\textit{(i) RTL Pre-processing}} that extracts structured representations of data and control dependencies; \ul{\textit{(ii) Divide and Conquer (Structural Breakdown)}}, which partitions the design into fine-grained blocks for modular analysis by LLM; \ul{\textit{(iii) LLM Engine}}, where the LLM evaluates data propagation at the block level; \ul{\textit{(iv) Response Integration}}, which synthesizes results into a comprehensive IFT assessment. Unlike traditional IFT, LLM-IFT dynamically infers security properties by LLM-based analyzing module dependencies, security-critical assets, inputs, outputs, and control paths, enhancing tracking across the SoC, reducing false positives and improving threat detection accuracy. The key contributions of this work are as follows:

\begin{itemize}[leftmargin=*]
    \item We introduce LLM-IFT, a new LLM-based framework for recursive IFT that dynamically verifies security properties.
    \item LLM-IFT establishes a structural breakdown and integration that enables IFT to scale for complex SoC designs.
    \item Tested on Trust-Hub \cite{salmani2013ondesign} and Hackathons benchmarks \cite{salmani2013ondesign}, LLM-IFT achieves 0\% false positive and 100\% accuracy.
\end{itemize}



\section{Related Works}

\subsection{Limitation of Hardware Security Verification Methods}

With the growing complexity of hardware designs and increasing reliance on 3PIPs, multiple approaches have been engaged over the years to validate RTL designs \cite{solt2022cellift}: 

\begin{itemize}[leftmargin=*]
    \item \textbf{Simulation-based testing} --- It relies on ad hoc test cases, making it error-prone, time-consuming, and unsuitable for continuous integration \cite{dessouky2019hardfails, rajendran2023hunter}. Its effectiveness is further limited by the difficulty of crafting comprehensive test scenarios that cover all potential vulnerabilities.
    \item \textbf{Fuzz and Penetration Testing} --- They leverage mutation-based test case generation and feedback-driven exploration based on security properties (security cost functions) \cite{hossain2023socfuzzer, al2023sharpen}. However, their low convergence rate makes them inefficient, especially at the scale of large SoC designs.
    \item \textbf{Assertion-based formal security verification} --- It defines security properties using SystemVerilog assertions (SVA) and employs formal verification methods to mathematically prove their correctness \cite{aftabjahani2021special}. Despite its precision, formal verification suffers from state explosion, making it impractical for complex designs and limiting its scalability.
    \item \textbf{ML-based validation} --- It is often relying on graph neural network (GNN)-based approach \cite{yasaei2022hardware, fan2024efficient}.  ML-based approaches heavily rely on labeled datasets, which are scarce in hardware security, lowering their effectiveness.  
\end{itemize}

\subsection{IFT in Hardware Security Verification}

IFT \cite{hu2021hardware} analyzes data propagation in circuits to detect unauthorized information flows. Gate-Level Information Flow Tracking (GLIFT), originally proposed by Tiwari et al \cite{tiwari2009complete}, applies taint labels to individual bits and tracks their propagation through shadow logic. Over the years, various enhancements to GLIFT have been proposed \cite{hu2016detecting, qin2019theorem, goli2019security, zhao2024static}. As an instance, Qin et al. \cite{qin2019theorem} integrate GLIFT with theorem proving in Coq, formalizing security properties and verifying Trojan conditions in OpenRISC and RSA cryptographic cores. Zhao et al. \cite{zhao2024static} extend GLIFT with bounded model checking (BMC), offering counterexamples when security policies are violated. While these gate-level IFTs provide precise tracing of unauthorized information flows, its fine-grained (gate-level) analysis leads to huge overhead, making it impractical to scale for large SoCs \cite{ardeshiricham2017register}. Hence, IFTs have evolved to operate at higher levels of abstraction, e.g., register-transfer level (RTL), earlier detection of security vulnerabilities during the design phase, reducing the complexity associated with low-level gate tracking. For instance, CellIFT \cite{solt2022cellift} injects taining logic at the coarse-grained logic netlist (at the macro level using Yosys \cite{wolf2013yosys}) to achieve scalable flow tracking. However, these techniques still incur substantial instrumentation costs due to the extensive tainting required for accurate tracking \cite{solt2024hybridift}. 

\subsection{LLM for Hardware Security Verification}

With the breakthroughs in LLMs and their widespread application, many recent studies have also demonstrated their role in security verification, including vulnerability detection in hardware designs \cite{akyash2024evolutionary}. While in software security the use of LLM for static taint analysis has been investigated \cite{li2024llm}, to the best of our knowledge, in hardware, LLMs being leveraged for two general purposes: (i) system-level security validation through property generation and code analysis (reasoning) \cite{paria2023divas, akyash2024self}, and (ii) SVA generation followed by formal verification \cite{orenes2023using, kande2024security}, enabling property-driven RTL security enforcement. However, no prior work has explored the use of LLMs for Hardware IFT, which has the potential to significantly reduce the cost of taint instrumentation by leveraging LLMs’ reasoning capabilities to infer security properties dynamically.

\section{Methodolody: LLM-IFT}

Alg. \ref{alg:top_view_steps} shows the top view of LLM-IFT, which consists of four main steps. The following provide details of each step. 

\subsection{Step \raisebox{.5pt}{\textcircled{\raisebox{-.9pt} {1}}} --- RTL Pre-processing}

The first step in LLM-IFT involves extracting the structure of the hardware design to enable IFT analysis. The hardware designs are parsed at the RTL level, extracting module definitions along with their hierarchical relationships, which are required for further divide and conquer in Step \raisebox{.5pt}{\textcircled{\raisebox{-.9pt} {2}}}. The parsed output is then processed to structure the module hierarchy, ensuring that master-slave (topological) relationships between modules are accurately captured. Since our approach operates at the RTL level, it falls under RTL-based IFT (RTLIFT).
\subsection{Step \raisebox{.5pt}{\textcircled{\raisebox{-.9pt} {2}}} --- Divide and Conquer (Structural Breakdown)}

Once the module hierarchy is established, a directed acyclic graph (DAG) is constructed w.r.t. the required granularity\footnote{LLM-IFT can define granularity to apply breakdown as needed for higher accurate reasoning provided by LLM. Here we use module-level, where nodes represent modules and directed edges represent inter-module dependencies.}. Hence, the entire hardware design is modeled as a DAG \( G = (M, E) \) where: \(M = M_1, M_2, ..., M_n\) represents the set of all modules in the design and \(E\) the set of directed edges that capture dependencies such that \((M_i, M_j) \in E\) if module \(M_j\) depends on module \(M_i\). After the extraction, to efficiently store the dependencies, an adjacency list \(A\) is constructed: \(A(M_i) = \{M_j | (M_i,M_j) \in  E\}\). 
To enable IFT automation using LLM, a topological sorting algorithm is applied to the DAG, yielding an ordered sequence of modules: \(\sigma = (M_{s_1}, M_{s_2}, ..., M_{s_n})\). This sorting ensures that all dependencies are resolved before a module is processed, which is essential for (sequential) flow tracking.
Each module \(M_i\) is assigned a hierarchy Level \(L(M_i)\), representing its depth w.r.t. the starting point for IFT analysis (w.r.t. the origin of security assets):
\(L(M_i) = 1 + \max (L(M_j)), \quad \forall M_j \in A(M_i)\). The extracted structure is then used to run the IFT using LLM. 

\subsection{Step \raisebox{.5pt}{\textcircled{\raisebox{-.9pt} {3}}} --- LLM Engine (Prompting)}

The LLM's initial prompt includes the sorted module list and adjacency matrix of dependencies. We also designed the prompt to process the upcoming modules. Each module \(M_i\) is analyzed sequentially based on the order defined by the topological sorting \(\sigma\) while using insights from previously analyzed modules. The relationships between modules allow the extraction of an ancestor set \(P(M_i)\) where: \(P(M_i) = \{ M_j \mid M_j \to M_i \text{ in } G \}\) which consists of all modules that influence the behavior of \(M_i\). The objective is to track both explicit and implicit data flows within \(M_i\) to determine whether security asset is propagated in an unauthorized manner.
For each module \(M_i\) an evaluation function \(f(P(M_i), D(M_i), A(M_i), T_i)\) is applied where \(D(M_i)\) represents the extracted Verilog representation of the module, \(A(M_i)\) denotes the list of ancestors, \(D(M_i)\) is the list of modules that are dependent on the module \(M_i\) and influenced by it and \(T_i\) refers to the selected IFT techniques, such as gate-level IFT and net-level IFT. The evaluation function generates a structured prompt for the LLM, including details about the module, its dependencies, and IFT techniques. The LLM processes this prompt and returns a structured output containing: \(\Theta(M_i) = \{ S_i, O_i, \Psi_i, \Lambda_i \}\) where \(S_i\) represents the sensitive data sources within the module, \(O_i\) denotes the assets influenced by the data flow, \(\Psi_i\) describes logical transformations applied to the data and \(\Lambda_i\) captures the flow of information between internal and external module components. The results of the analysis are stored in an accumulated context \(C\) which retains findings from previously analyzed modules: \(C = \bigcup_{k=1}^{i-1} f(M_k, P_k, D_k, A_k, T_k)\) ensuring that findings from previous modules contribute to subsequent analyses.
The evaluation process is iterative, in which the insights of each module are continuously refined and stored, allowing subsequent modules to benefit from earlier findings. This approach ensures that information propagation across interconnected modules is consistently tracked. 

\begin{algorithm}[t]
\scriptsize
\caption{Main Steps of LLM-IFT}
\label{alg:top_view_steps}
\begin{algorithmic}[1]
\Require Verilog design files
\Ensure Leakage detection and structured output report

\textbf{\ul{Step 1: Pre-processing \& Step 2: Structural breakdown}}
\State Parse Verilog files and extract module dependencies
\State Construct DAG representation \( G = (M, E) \) and adjacency list \( A \)
\State Perform topological sorting to obtain order \( \sigma = (M_{s_1}, M_{s_2}, ..., M_{s_n}) \)

\textbf{\ul{Step 3: LLM Engine (Prompting for IFT)}}
\For{each module \( M_i \) in \( \sigma \)}
    \State Retrieve ancestor modules \( P_i \) using \( A \)
    \State Extract Verilog code representation \( D_i \)
    \State Formulate structured prompt with \( M_i, P_i, D_i, A, T_i \) and submit to LLM
    \State LLM returns \( \Theta(M_i) = \{ S_i, O_i, \Psi_i, \Lambda_i \} \)
    \State Update accumulated context \( C = C \cup \Theta(M_i) \)
\EndFor

\textbf{\ul{Step 4: Response Integration (End-to-End Leakage Path Analysis)}}
\State Compute global assessment function \( g(C, G) \)
\If{\( g(C, G) = 1 \)}
    \State Construct leakage path sequence \( \Lambda = (M_{\lambda_1} \to M_{\lambda_2} \to ... \to M_{\lambda_k}) \)
    \State Identify transformations and sensitive flows \( \Psi = \{(M_i, s_i, o_i, M_j) \} \)
    \State Generate structured report \( \Omega = (V, M_v, \Lambda, T, E) \)
\Else
    \State No confirmed leakage detected
\EndIf

\State Output final structured JSON report
\end{algorithmic}
\end{algorithm}

\subsection{Step \raisebox{.5pt}{\textcircled{\raisebox{-.9pt} {4}}} --- Response Integration}

Once all modules have been assessed, a final analysis is performed, consolidating the accumulated insights to determine if an overall leakage path exists. Using the accumulated findings $C$, a structured prompt is formulated and submitted to the LLM which integrates the topologically sorted list of modules, the adjacency list representation, the previously collected IFT context detailing data transformation and the specified IFT techniques. The LLM processes this and returns an analysis of potential data leakage, providing an ordered sequence of information propagation across modules. If leakage is detected, the LLM returns a detailed stepwise sequence of the whole leakage. Formally, the LLM returns a structured JSON output \(\Omega\) detailing \(\Omega = (V, M_v, \Lambda, Z, E)\) where \(V\) is the boolean flag mentioning the vulnerability presence, \(Mv\) represents the vulnerable modules, \(\Lambda\) represents the leakage path, \(Z\) represents the type of leakage and \(E\) provides the detailed explanation of how the leak occurs.

\section{Results and Evaluation}

\subsection{Dataset and Experimental Setup}

\begin{table}[b]
\scriptsize
\centering
\caption{Dataset Overview used for Exploration of LLM-IFT}
\label{tab:dataset_overview}
\begin{tabular}{@{} l *{5}c @{}}
\toprule 
\textbf{Category} & \textbf{No. of Designs} \\
\cmidrule(r){1-1}\cmidrule(l){2-2}
Trust-Hub - HT (Leakage-affected, Leakage-free)  & 14, 14 \\ 
\cmidrule(r){1-1}\cmidrule(l){2-2}
OpenTitan - Hackatons (Leakage-affected, Leakage-free) & 3, 2 \\
\bottomrule
\end{tabular}
\end{table}

To evaluate LLM-IFT, two sets of data have been used: (i) 14 hardware (in RTL) designs are derived from Trust-hub \cite{salmani2013ondesign} as shown in Table \ref{tab:dataset_overview}\footnote{We focused on a Trust-hub dataset consisting of 212 designs with Trojan implementations. Among them, 14 are designed to raise information leakage vulnerabilities (desired cases for the IFT for security).}; (ii) The OpenTitan SoC \cite{ahmad2023fixing} (in RTL) consisting of 3 injected buggy modules known to exhibit potential information flow related vulnerabilities and two non-buggy modules serving as the baseline.   Additionally, a subset of the database equipped with code-division multiple access (CDMA)-based implementation of vulnerabilities where partial leaks occur over multiple clock cycles, challenging LLM-IFT’s ability to track dynamic, time-distributed information flows. The implementation involved multiple tools: Yosys was used for synthesis and module hierarchy extraction, a custom script for graph construction, and NetworkX for building dependency graphs and adjacency lists. GPT-4o, DeepSeek V3, Qwen Plus and Llama 3.3 powered the LLM-based reasoning for IFT analysis, with LangChain managing context throughout the process. LLM-IFT combines two IFT techniques (\(T_i\)): gate-level, tracking logic propagation through gates, and net-level, analyzing signal propagation through wires\footnote{The definitions of gate-level/net-level IFT are provided to the LLMs so that data and control flow within hardware designs can be analyzed accordingly.}.

Table \ref{tab:results_with_n_without_modular} reflects the success rate of LLM-IFT, before and after enabling the concept of divide and conquer (structural breakdown) for running the IFT using LLM. As seen, without this concept, the model (GPT-4o) achieved 64.9\% accuracy, while it faces a very high false positive rate. This is while by introducing modular analysis incorporating hierarchical dependencies, the accuracy went up to 100\%, while the false positive rate is minimized. Some of our key observations in terms of structural breakdown are: (i) The LLM was able to better contextualize information flow and module interactions; (ii) LLMs showed limited sensitivity in detecting covert leakage when no breakdown is used; (iii) False negatives were observed in cases where the leakage covered multiple module interactions, highlighting the limitations of LLM in maintaining a long-range hierarchical context.

\begin{table}[t]
\scriptsize
\centering
\caption{LLM-IFT Detection over Leakage-based Vulnerabilities.}
\label{tab:results_with_n_without_modular}
\setlength\tabcolsep{4pt}
\begin{tabular}{@{} l *{5}c @{}}
\toprule 
\textbf{LLM Models} & \textbf{Approach} & \textbf{Success Rate} & \textbf{False Positive Rate$^*$} \\
\cmidrule(r){1-1}\cmidrule(r){2-2}\cmidrule(r){3-3}\cmidrule(l){4-4}
\multirow{2}{*}{GPT-4o} & w/o divide and conquer & 64.29\% & 71.40\% \\
                        & \textbf{with divide and conquer} & \textbf{100\%} & \textbf{0\%} \\
\cmidrule(r){1-1}\cmidrule(r){2-2}\cmidrule(r){3-3}\cmidrule(l){4-4}
\multirow{2}{*}{DeepSeek V3} & w/o divide and conquer & 62.50\%  & 75\% \\
                             & \textbf{with divide and conquer} & \textbf{100\%} & \textbf{0\%} \\
\cmidrule(r){1-1}\cmidrule(r){2-2}\cmidrule(r){3-3}\cmidrule(l){4-4}
\multirow{2}{*}{Qwen Plus} & w/o divide and conquer & 68.75\%  & 62.5\%  \\
                             & \textbf{with divide and conquer} & \textbf{81.25\%} & \textbf{25\%}  \\
\cmidrule(r){1-1}\cmidrule(r){2-2}\cmidrule(r){3-3}\cmidrule(l){4-4}
\multirow{2}{*}{Llama 3.3} & w/o divide and conquer & 50\%  & 100\%  \\
                             & \textbf{with divide and conquer} & \textbf{81.25\%}  & \textbf{37.5\%} \\
\bottomrule
\multicolumn{4}{l}{$^*$: Calculated by running LLM-IFT on 16 benchmarks—8 leakage-free, 8 with leakage.} \\

\end{tabular}
\end{table}

\subsection{Showcase 1: AES Key Leakage Detection using LLM-IFT}

\begin{table}[t]
\scriptsize
\centering
\setlength\tabcolsep{1.5pt}
\caption{Propagation Leading to AES Leakage Extracted by LLM-IFT.}
\label{tab:propagation_aes}
\begin{tabular}{@{} l *{5}l @{}}
\toprule 
\textbf{Step} & \textbf{Description} \\
\cmidrule(r){1-1}\cmidrule(l){2-2}
\multirow{2}{*}{Initial step} & \texttt{key[7:0] = 8'b10101010} \\
 & \texttt{data[19:0] = 20'hABCDE} \\
\cmidrule(r){1-1}\cmidrule(l){2-2}
\multirow{2}{*}{LFSR seeding \& extract} & \textbf{On reset}, \texttt{lfsr\_stream = 20'hABCDE} \\
 & \texttt{lfsr[7:0] = 11011110}  \\
\cmidrule(r){1-1}\cmidrule(l){2-2}
\multirow{4}{*}{Trojan's XOR modulation} & \texttt{load[0-7] = key[0] $\oplus$ lfsr[0] = 00000000} \\
 & \texttt{$\vdots$} \\
 & \texttt{load[48-55] = key[6] $\oplus$ lfsr[6] = 11111111} \\
 & \texttt{load[56-63] = key[7] $\oplus$ lfsr[7] = 00000000} \\
\cmidrule(r){1-1}\cmidrule(l){2-2}
Expose to Load & \texttt{0x00 00 FF 00 FF FF FF 00} \\
Load to Capacitance & \texttt{Capacitance = 0x00 00 FF 00 FF FF FF 00} \\
\multirow{2}{*}{Extract Key} & \texttt{lfsr $\oplus$ capacitance = key} \\ 
& \texttt{01111011 $\oplus$ 00101110 = 01010101 (key)} \\
\bottomrule
\end{tabular}
\end{table}

\begin{figure}[b]
    \centering
    \includegraphics[width=1.0\linewidth]{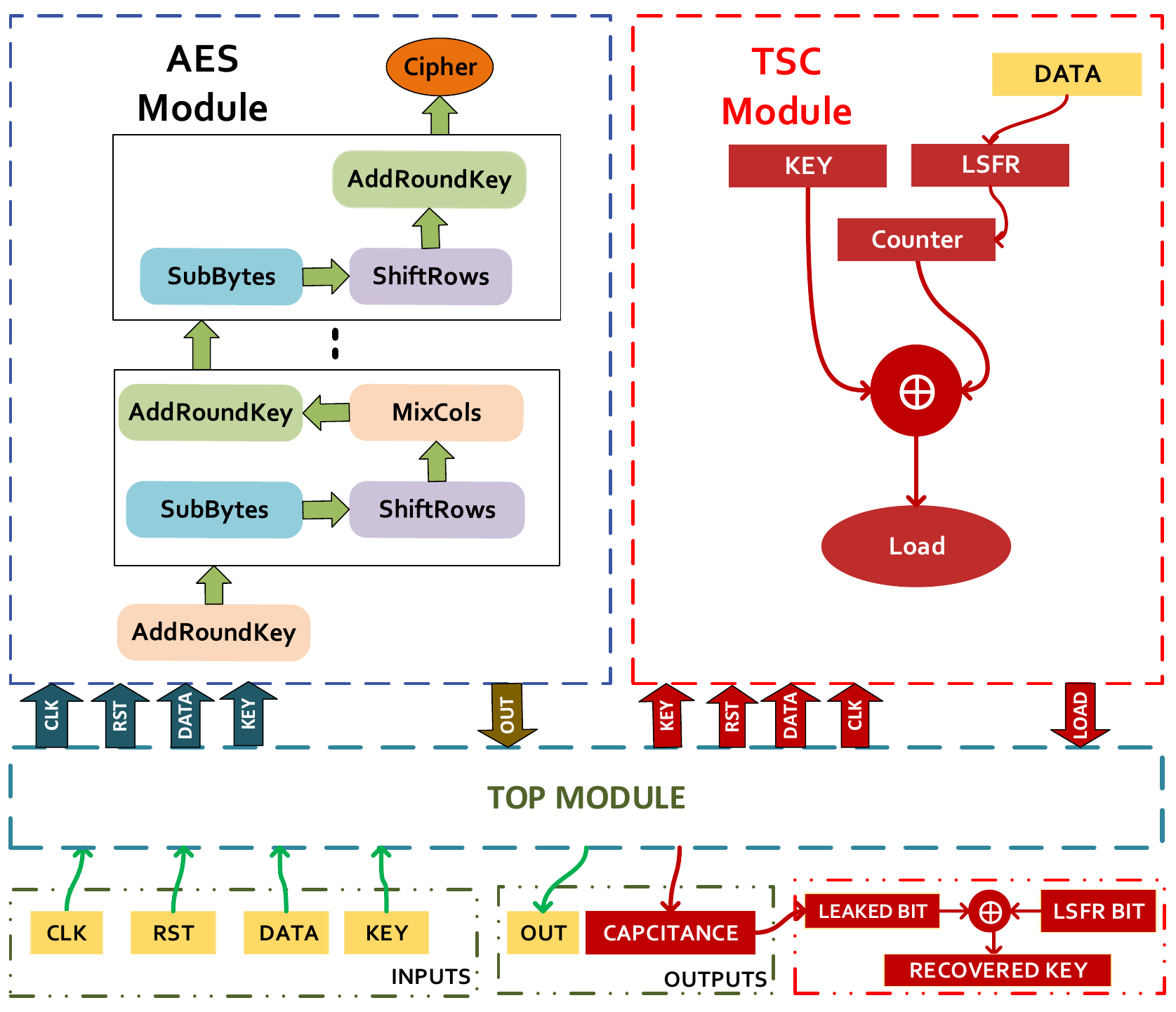}
    \caption{AES Module from Trust-Hub w/ HT and Leakage Path}
    \label{fig:aes_leak_path}
\end{figure}

From Fig. \ref{fig:aes_leak_path}, we can see the whole path for the side channel leakage. The leakage path begins at the top module. It receives two critical inputs: the plaintext (\texttt{DATA}) and the encryption key (\texttt{KEY}). The expected output of the encryption process is the ciphertext (\texttt{OUT}), while an unintended capacitance output is exploited to leak information. The TSC module, which houses the Trojan, extracts and modulates the key with a PRNG-generated sequence. The LFSR counter, seeded with plaintext, spreads the leakage over time. The Trojan logic XORs key bits with LFSR output. The manipulated key bits are passed through the load signal of TSC, which is mapped to the top module capacitance. The attacker knowing the PRNG sequence, reconstructs the key by XORing observed data. 

\begin{table}[t]
\scriptsize
\centering
\caption{Extracted Propagation Path for SoC Leakage Path}
\renewcommand{\arraystretch}{1.3}
\begin{tabular}{@{} l *{5}l @{}}
\toprule
\textbf{Step} & \textbf{Description} \\
\cmidrule(r){1-1}\cmidrule(l){2-2}
Input Data from memory & \texttt{config\_mem\_data} = 32'h0badc0de \\
\cmidrule(r){1-1}\cmidrule(l){2-2}
Step 1 & \texttt{config\_mem\_data} $\rightarrow$ \texttt{config\_mem\_unit.dout} \\
\cmidrule(r){1-1}\cmidrule(l){2-2}
Step 2 & \texttt{config\_mem\_unit.dout} $\rightarrow$\\ & \texttt{status\_transmitter\_unit.config\_in} \\
\cmidrule(r){1-1}\cmidrule(l){2-2}
Step 3 & \texttt{config\_in} compared with 32'hdeadc0de \\
\cmidrule(r){1-1}\cmidrule(l){2-2}
Step 4 & If condition met,\\ & LSB of \texttt{config\_in} $\rightarrow$ \texttt{tx\_signal} \\
\cmidrule(r){1-1}\cmidrule(l){2-2}
Step 5 & \texttt{tx\_signal} $\rightarrow$ \\ & \texttt{soc\_integration\_top.final\_tx} \\
\cmidrule(r){1-1}\cmidrule(l){2-2}
Exposed Output & \texttt{final\_tx} transmits extracted bit externally \\
\bottomrule
\end{tabular}
\label{tab:soc_leakage_path}
\end{table}

\subsection{Showcase 2: SoC Information Leakage Analysis}

To evaluate the scalability of LLM-IFT for detecting information leakage at the SoC level, we applied it to various modules of the bug-injected OpenTitan SoC \cite{ahmad2023fixing}. Using the sequential hierarchical divide-and-conquer in LLM-IFT, it systematically decomposes the data flow across various design levels (IPs). By tracking data dependencies in a structured manner, then LLM-IFT detects vulnerabilities through step-by-step propagation through interconnected modules. As an instance, as shown in Table \ref{tab:soc_leakage_path}, a sensitive hard-coded valu initially stored in a memory unit may undergo a series of transformations before reaching an external interface. As shown, LLM-IFT isolates critical transition points, such as conditional operations or signal mappings, that contribute to unintended exposure. This hierarchical approach ensures that even in complex SoC designs, LLM-IFT maintains efficiency and accuracy, successfully scaling to large hardware systems by methodically breaking down the verification process.


\section{Conclusion and Future Works}

This paper introduces LLM-IFT, an LLM-driven information flow tracking (IFT) scheme for information leakage detection in hardware designs. LLM-IFT, consisting of four main phases, i.e., RTL structural pre-processing, structured design extraction, module-wise IFT analysis, and final leakage assessment, identifies unauthorized information flows across interconnected modules in hardware design at various sizes and complexities. With module-wise IFT analysis, LLM-IFT shows a 100\% success rate over Trust-hub vulnerability-injected cases (information leakage due to hardware Trojan), while the false positive rate is minimized, compared to the traditional form of LLM utilization. Future work will focus on enhancing context management, coupled with formal verification techniques, and fine-tuning LLMs on hardware security datasets to further improve the accuracy and automation of IFT in real-world SoC security verification.

\bibliographystyle{IEEEtran}
\bibliography{refs}

\begin{thebibliography}{10}
\providecommand{\url}[1]{#1}
\csname url@samestyle\endcsname
\providecommand{\newblock}{\relax}
\providecommand{\bibinfo}[2]{#2}
\providecommand{\BIBentrySTDinterwordspacing}{\spaceskip=0pt\relax}
\providecommand{\BIBentryALTinterwordstretchfactor}{4}
\providecommand{\BIBentryALTinterwordspacing}{\spaceskip=\fontdimen2\font plus
\BIBentryALTinterwordstretchfactor\fontdimen3\font minus \fontdimen4\font\relax}
\providecommand{\BIBforeignlanguage}[2]{{%
\expandafter\ifx\csname l@#1\endcsname\relax
\typeout{** WARNING: IEEEtran.bst: No hyphenation pattern has been}%
\typeout{** loaded for the language `#1'. Using the pattern for}%
\typeout{** the default language instead.}%
\else
\language=\csname l@#1\endcsname
\fi
#2}}
\providecommand{\BIBdecl}{\relax}
\BIBdecl

\bibitem{tehranipoor2024hardware}
M.~Tehranipoor, K.~Z. Azar, N.~Asadizanjani, F.~Rahman, H.~M. Kamali, and F.~Farahmandi, \emph{Hardware Security}.\hskip 1em plus 0.5em minus 0.4em\relax Springer, 2024.

\bibitem{hu2020overview}
W.~Hu, C.-H. Chang, A.~Sengupta, S.~Bhunia, R.~Kastner, and H.~Li, ``An overview of hardware security and trust: Threats, countermeasures, and design tools,'' \emph{IEEE Transactions on Computer-Aided Design of Integrated Circuits and Systems}, vol.~40, no.~6, pp. 1010--1038, 2020.

\bibitem{farzana2019soc}
N.~Farzana, F.~Rahman, M.~Tehranipoor, and F.~Farahmandi, ``Soc security verification using property checking,'' in \emph{2019 IEEE International Test Conference (ITC)}.\hskip 1em plus 0.5em minus 0.4em\relax IEEE, 2019, pp. 1--10.

\bibitem{azar2022fuzz}
K.~Z. Azar, M.~M. Hossain, A.~Vafaei, H.~Al~Shaikh, N.~N. Mondol, F.~Rahman, M.~Tehranipoor, and F.~Farahmandi, ``Fuzz, penetration, and ai testing for soc security verification: Challenges and solutions,'' \emph{Cryptology ePrint Archive}, 2022.

\bibitem{witharana2022survey}
H.~Witharana, Y.~Lyu, S.~Charles, and P.~Mishra, ``A survey on assertion-based hardware verification,'' \emph{ACM Computing Surveys (CSUR)}, vol.~54, no. 11s, pp. 1--33, 2022.

\bibitem{jain2021survey}
A.~Jain, Z.~Zhou, and U.~Guin, ``Survey of recent developments for hardware trojan detection,'' in \emph{2021 ieee international symposium on circuits and systems (iscas)}.\hskip 1em plus 0.5em minus 0.4em\relax IEEE, 2021, pp. 1--5.

\bibitem{hossain2024fuzzing}
M.~M. Hossain, K.~Z. Azar, F.~Rahman, F.~Farahmandi, and M.~Tehranipoor, ``Fuzzing for automated soc security verification: Challenges and solution,'' \emph{IEEE Design \& Test}, 2024.

\bibitem{fan2024efficient}
R.~Fan, Y.~Tang, H.~Sun, J.~Liu, and H.~Li, ``An efficient ml-based hardware trojan localization framework for rtl security analysis,'' in \emph{Proceedings of the 2024 ACM/IEEE International Symposium on Machine Learning for CAD}, 2024, pp. 1--7.

\bibitem{yasaei2022hardware}
R.~Yasaei, L.~Chen, S.-Y. Yu, and M.~A. Al~Faruque, ``Hardware trojan detection using graph neural networks,'' \emph{IEEE Transactions on Computer-Aided Design of Integrated Circuits and Systems}, 2022.

\bibitem{hu2021hardware}
W.~Hu, A.~Ardeshiricham, and R.~Kastner, ``Hardware information flow tracking,'' \emph{ACM Computing Surveys (CSUR)}, vol.~54, no.~4, pp. 1--39, 2021.

\bibitem{tai2020multi}
Y.~Tai, W.~Hu, L.~Zhang, D.~Mu, and R.~Kastner, ``A multi-flow information flow tracking approach for proving quantitative hardware security properties,'' \emph{Tsinghua Science and Technology}, vol.~26, no.~1, pp. 62--71, 2020.

\bibitem{nahiyan2017hardware}
A.~Nahiyan, M.~Sadi, R.~Vittal, G.~Contreras, D.~Forte, and M.~Tehranipoor, ``Hardware trojan detection through information flow security verification,'' in \emph{2017 IEEE International Test Conference (ITC)}.\hskip 1em plus 0.5em minus 0.4em\relax IEEE, 2017, pp. 1--10.

\bibitem{reimann2021qflow}
L.~M. Reimann, L.~Hanel, D.~Sisejkovic, F.~Merchant, and R.~Leupers, ``Qflow: Quantitative information flow for security-aware hardware design in verilog,'' in \emph{2021 IEEE 39th International Conference on Computer Design (ICCD)}.\hskip 1em plus 0.5em minus 0.4em\relax IEEE, 2021, pp. 603--607.

\bibitem{solt2022cellift}
F.~Solt, B.~Gras, and K.~Razavi, ``$\{$CellIFT$\}$: Leveraging cells for scalable and precise dynamic information flow tracking in $\{$RTL$\}$,'' in \emph{31st USENIX Security Symposium (USENIX Security 22)}, 2022, pp. 2549--2566.

\bibitem{solt2024hybridift}
F.~Solt and K.~Razavi, ``Hybridift: Scalable memory-aware dynamic information flow tracking for hardware,'' in \emph{ICCAD (International Conference on Computer-Aided Design)}, 2024.

\bibitem{jin2024large}
B.~Jin, G.~Liu, C.~Han, M.~Jiang, H.~Ji, and J.~Han, ``Large language models on graphs: A comprehensive survey,'' \emph{IEEE Transactions on Knowledge and Data Engineering}, 2024.

\bibitem{salmani2013ondesign}
H.~Salmani, M.~Tehranipoor, and R.~Karri, ``On design vulnerability analysis and trust benchmarks development,'' in \emph{2013 IEEE 31st international conference on computer design (ICCD)}.\hskip 1em plus 0.5em minus 0.4em\relax IEEE, 2013, pp. 471--474.

\bibitem{dessouky2019hardfails}
G.~Dessouky, D.~Gens, P.~Haney, G.~Persyn, A.~Kanuparthi, H.~Khattri, J.~M. Fung, A.-R. Sadeghi, and J.~Rajendran, ``$\{$HardFails$\}$: insights into $\{$software-exploitable$\}$ hardware bugs,'' in \emph{28th USENIX Security Symposium (USENIX Security 19)}, 2019, pp. 213--230.

\bibitem{rajendran2023hunter}
S.~R. Rajendran, S.~Tarek, B.~M. Hicks, H.~M. Kamali, F.~Farahmandi, and M.~Tehranipoor, ``Hunter: Hardware underneath trigger for exploiting soc-level vulnerabilities,'' in \emph{2023 Design, Automation \& Test in Europe Conference \& Exhibition (DATE)}.\hskip 1em plus 0.5em minus 0.4em\relax IEEE, 2023, pp. 1--6.

\bibitem{hossain2023socfuzzer}
M.~M. Hossain, A.~Vafaei, K.~Z. Azar, F.~Rahman, F.~Farahmandi, and M.~Tehranipoor, ``Socfuzzer: Soc vulnerability detection using cost function enabled fuzz testing,'' in \emph{2023 Design, Automation \& Test in Europe Conference \& Exhibition (DATE)}.\hskip 1em plus 0.5em minus 0.4em\relax IEEE, 2023, pp. 1--6.

\bibitem{al2023sharpen}
H.~Al-Shaikh, A.~Vafaei, M.~M.~M. Rahman, K.~Z. Azar, F.~Rahman, F.~Farahmandi, and M.~Tehranipoor, ``Sharpen: Soc security verification by hardware penetration test,'' in \emph{Proceedings of the 28th Asia and South Pacific Design Automation Conference}, 2023, pp. 579--584.

\bibitem{aftabjahani2021special}
S.~Aftabjahani, R.~Kastner, M.~Tehranipoor, F.~Farahmandi, J.~Oberg, A.~Nordstrom, N.~Fern, and A.~Althoff, ``Special session: Cad for hardware security-automation is key to adoption of solutions,'' in \emph{2021 IEEE 39th VLSI Test Symposium (VTS)}.\hskip 1em plus 0.5em minus 0.4em\relax IEEE, 2021, pp. 1--10.

\bibitem{tiwari2009complete}
M.~Tiwari, H.~M. Wassel, B.~Mazloom, S.~Mysore, F.~T. Chong, and T.~Sherwood, ``Complete information flow tracking from the gates up,'' in \emph{Proceedings of the 14th international conference on Architectural support for programming languages and operating systems}, 2009, pp. 109--120.

\bibitem{hu2016detecting}
W.~Hu, B.~Mao, J.~Oberg, and R.~Kastner, ``Detecting hardware trojans with gate-level information-flow tracking,'' \emph{Computer}, vol.~49, no.~8, pp. 44--52, 2016.

\bibitem{qin2019theorem}
M.~Qin, W.~Hu, X.~Wang, D.~Mu, and B.~Mao, ``Theorem proof based gate level information flow tracking for hardware security verification,'' \emph{Computers \& Security}, vol.~85, pp. 225--239, 2019.

\bibitem{goli2019security}
M.~Goli, M.~Hassan, D.~Gro{\ss}e, and R.~Drechsler, ``Security validation of vp-based socs using dynamic information flow tracking,'' \emph{it-Information Technology}, vol.~61, no.~1, pp. 45--58, 2019.

\bibitem{zhao2024static}
Y.~Zhao, G.~Qu, Q.~Zhang, Y.~Li, Z.~Li, and J.~He, ``Static gate-level information flow for hardware information security with bounded model checking,'' in \emph{2024 IEEE 42nd VLSI Test Symposium (VTS)}.\hskip 1em plus 0.5em minus 0.4em\relax IEEE, 2024, pp. 1--7.

\bibitem{ardeshiricham2017register}
A.~Ardeshiricham, W.~Hu, J.~Marxen, and R.~Kastner, ``Register transfer level information flow tracking for provably secure hardware design,'' in \emph{Design, Automation \& Test in Europe Conference \& Exhibition (DATE), 2017}.\hskip 1em plus 0.5em minus 0.4em\relax IEEE, 2017, pp. 1691--1696.

\bibitem{wolf2013yosys}
C.~Wolf, J.~Glaser, and J.~Kepler, ``Yosys-a free verilog synthesis suite,'' in \emph{Proceedings of the 21st Austrian Workshop on Microelectronics (Austrochip)}, vol.~97, 2013.

\bibitem{akyash2024evolutionary}
M.~Akyash and H.~M~Kamali, ``Evolutionary large language models for hardware security: A comparative survey,'' in \emph{Proceedings of the great lakes symposium on VLSI 2024}, 2024, pp. 496--501.

\bibitem{li2024llm}
Z.~Li, S.~Dutta, and M.~Naik, ``Llm-assisted static analysis for detecting security vulnerabilities,'' \emph{arXiv preprint arXiv:2405.17238}, 2024.

\bibitem{paria2023divas}
S.~Paria, A.~Dasgupta, and S.~Bhunia, ``Divas: An llm-based end-to-end framework for soc security analysis and policy-based protection,'' \emph{arXiv preprint arXiv:2308.06932}, 2023.

\bibitem{akyash2024self}
M.~Akyash and H.~M. Kamali, ``Self-hwdebug: Automation of llm self-instructing for hardware security verification,'' in \emph{2024 IEEE Computer Society Annual Symposium on VLSI (ISVLSI)}.\hskip 1em plus 0.5em minus 0.4em\relax IEEE, 2024, pp. 391--396.

\bibitem{orenes2023using}
M.~Orenes-Vera, M.~Martonosi, and D.~Wentzlaff, ``Using llms to facilitate formal verification of rtl,'' \emph{arXiv preprint arXiv:2309.09437}, 2023.

\bibitem{kande2024security}
R.~Kande, H.~Pearce, B.~Tan, B.~Dolan-Gavitt, S.~Thakur, R.~Karri, and J.~Rajendran, ``(security) assertions by large language models,'' \emph{IEEE Transactions on Information Forensics and Security}, 2024.

\bibitem{ahmad2023fixing}
B.~Ahmad, S.~Thakur, B.~Tan, R.~Karri, and H.~Pearce, ``Fixing hardware security bugs with large language models,'' \emph{arXiv preprint arXiv:2302.01215}, 2023.

\end{thebibliography}

\end{document}